
\input harvmac.tex
\Title{CTP/TAMU-44/92}{{Exact Solutions in Bosonic and Heterotic String Theory}
\footnote{$^\dagger$}{Work supported in part by NSF grant PHY-9106593.}}

\centerline{
Ramzi~R.~ Khuri\footnote{$^*$}{Supported by a World Laboratory Fellowship.}}
\bigskip\centerline{Center for Theoretical Physics}
\centerline{Texas A\&M University}\centerline{College Station, TX 77843}

\vskip .3in
We review some exact solitonic solutions of string theory with higher-membrane
structure. These include an axionic instanton solution of bosonic string
theory as well as multi-instanton and multimonopole solutions of heterotic
string theory. The heterotic solutions reveal some interesting aspects of
string theory as a theory of quantum gravity.

\Date{5/92}
\def\wbg{\overline{\beta}^G_{\mu\nu}}
\def\wbb{\overline{\beta}^B_{\mu\nu}}
\def\wbd{\overline{\beta}^\Phi}

\def\sqr#1#2{{\vbox{\hrule height.#2pt\hbox{\vrule width
.#2pt height#1pt \kern#1pt\vrule width.#2pt}\hrule height.#2pt}}}
\def\Box{\mathchoice\sqr64\sqr64\sqr{4.2}3\sqr33}
\def\rijkl{R^i{}_{jkl}}
\def\grijkl{\hat R^i{}_{jkl}}

\def\met {g_{\mu\nu}}

\lref\reyone {S.--J.~Rey {\it Axionic String Instantons
and their Low-Energy Implications}, Proceedings, Tuscaloosa 1989,
Superstrings and particle theory, p.291.}

\lref\reytwo {S.--J.~Rey, Phys. Rev. {\bf D43} (1991) 526.}

\lref\abenone{I.~Antoniadis, C.~Bachas, J.~Ellis and D.~V.~Nanopoulos,
Phys. Lett. {\bf B211} (1988) 393.}

\lref\abentwo{I.~Antoniadis, C.~Bachas, J.~Ellis and D.~V.~Nanopoulos,
Nucl. Phys. {\bf B328} (1989) 117.}

\lref\mtone{R.~R.~Metsaev and A.~A.~Tseytlin, Phys. Lett.
{\bf B191} (1987) 354.}

\lref\mttwo{R.~R.~Metsaev and A.~A.~Tseytlin,
Nucl. Phys. {\bf B293} (1987) 385.}

\lref\cfmp{C.~G.~Callan, D.~Friedan, E.~J.~Martinec
and M.~J.~Perry, Nucl. Phys. {\bf B262} (1985) 593.}

\lref\ckp{C.~G.~Callan,
I.~R.~Klebanov and M.~J.~Perry, Nucl. Phys. {\bf B278} (1986) 78.}

\lref\love{C.~Lovelace, Phys. Lett. {\bf B135} (1984) 75.}

\lref\fridven{B.~E.~Fridling and A.~E.~M.~Van de Ven,
Nucl. Phys. {\bf B268} (1986) 719.}

\lref\gepwit{D.~Gepner and E.~Witten, Nucl. Phys. {\bf B278} (1986) 493.}

\lref\quartet{D.~J.~Gross,
J.~A.~Harvey, E.~J.~Martinec and R.~Rohm, Nucl. Phys. {\bf B256} (1985) 253.}

\lref\dine{M.~Dine, Lectures delivered at
TASI 1988, Brown University (1988) 653.}

\lref\brone{E.~A.~Bergshoeff and M.~de Roo, Nucl.
Phys. {\bf B328} (1989) 439.}

\lref\brtwo{E.~A.~Bergshoeff and M.~de Roo, Phys. Lett. {\bf B218} (1989) 210.}

\lref\chsone{C.~G.~Callan, J.~A.~Harvey and A.~Strominger, Nucl. Phys.
{\bf B359} (1991) 611.}

\lref\chstwo{C.~G.~Callan, J.~A.~Harvey and A.~Strominger, Nucl. Phys.
{\bf B367} (1991) 60.}

\lref\bpst{A.~A.~Belavin, A.~M.~Polyakov, A.~S.~Schwartz and Yu.~S.~Tyupkin,
Phys. Lett. {\bf B59} (1975) 85.}

\lref\thooft{G.~'t~Hooft, Nucl. Phys. {\bf B79} (1974) 276.}

\lref\hoofan{G.~'t~Hooft, Phys. Rev. Lett., {\bf 37} (1976) 8.}

\lref\wil{F.~Wilczek, in {\it Quark confinement and field theory},
Ed. D.~Stump and D.~Weingarten, John Wiley and Sons, New York (1977).}

\lref\cofa{E.~Corrigan and D.~B.~Fairlie, Phys. Lett. {\bf B67} (1977) 69.}

\lref\jackone{R.~Jackiw, C.~Nohl and C.~Rebbi, Phys. Rev. {\bf D15} (1977)
1642.}

\lref\jacktwo{R.~Jackiw, C.~Nohl and C.~Rebbi, in {\it Particles and
Fields}, Ed. David Boal and A.~N.~Kamal, Plenum Publishing Co., New York
(1978), p.199.}

\lref\rkinst{R.~R.~Khuri, Phys. Lett. {\bf B259} (1991) 261.}

\lref\rkscat{C.~G.~Callan and R.~R.~Khuri, Phys. Lett. {\bf B261} (1991) 363.}

\lref\rkmant{R.~R.~Khuri, {\it Manton Scattering of String Solitons}
PUPT-1270 (to appear in Nucl. Phys. {\bf B}).}

\lref\rkdg{R.~R.~Khuri, {\it Some Instanton Solutions in String
Theory}, Proceedings of the XXth International Conference on
Differential Geometric Methods in Theoretical Physics, World Scientific,
October 1991.}

\lref\rkthes{R.~R.~Khuri, {\it Solitons and Instantons in String Theory},
 Princeton University Doctoral Thesis, August 1991.}

\lref\rksing{M.~J.~Duff, R.~R.~Khuri and J.~X.~Lu, {\it String and
Fivebrane Solitons: Singular or Non-singular?}, Texas A\&M preprint,
CTP/TAMU-89/91 (to appear in Nucl. Phys. {\bf B}).}

\lref\rkorb{R.~R.~Khuri and H.~S.~La, {\it Orbits of a String around a
Fivebrane}, Texas A\&M preprint, CTP/TAMU-95/91
(submitted to Phys. Rev. Lett.).}

\lref\rkmot{R.~R.~Khuri and H.~S.~La, {\it String Motion in Fivebrane
Geometry}, Texas A\&M preprint, CTP/TAMU-98/91 (submitted to Nucl. Phys. B).}

\lref\rkmono{R.~R.~Khuri {\it A Multimonopole Solution in String Theory},
Texas A\&M preprint, CTP/TAMU-33/92 (submitted to Phys. Lett. B).}

\lref\rkmscat{R.~R.~Khuri {\it Scattering of String Monopoles},
Texas A\&M preprint, CTP/TAMU-34/92 (submitted to Phys. Lett. B).}

\lref\rkmonex{R.~R.~Khuri {\it A Heterotic Multimonopole Solution},
Texas A\&M preprint, CTP/TAMU-35/92 (submitted to Nucl. Phys. B).}

\lref\rkmonin{R.~R.~Khuri {\it Monopoles and Instantons in String Theory},
Texas A\&M preprint, CTP/TAMU-38/92 (submitted to Phys. Rev. D).}

\lref\rkexact{R.~R.~Khuri {\it Exact Solutions in Bosonic and Heterotic
String Theory}, Texas A\&M preprint, CTP/TAMU-35/92
(submitted to ).}

\lref\ginsp{P.~Ginsparg, Lectures delivered at
Les Houches summer session, June 28--August 5, 1988.}

\lref\swzw {W.~Boucher, D.~Friedan and A.~Kent, Phys. Lett.
{\bf B172} (1986) 316.}

\lref\dghrr{A.~Dabholkar, G.~Gibbons, J.~A.~Harvey and F.~Ruiz Ruiz,
Nucl. Phys. {\bf B340} (1990) 33.}

\lref\dabhar{A.~Dabholkar and J.~A.~Harvey,
Phys. Rev. Lett. {\bf 63} (1989) 478.}

\lref\prso{M.~K.~Prasad and C.~M.~Sommerfield, Phys. Rev. Lett. {\bf 35}
(1975) 760.}

\lref\jim{J.~A.~Harvey and J.~Liu, Phys. Lett. {\bf B268} (1991) 40.}

\lref\mantone{N.~S.~Manton, Nucl. Phys. {\bf B126} (1977) 525.}

\lref\manttwo{N.~S.~Manton, Phys. Lett. {\bf B110} (1982) 54.}

\lref\mantthree{N.~S.~Manton, Phys. Lett. {\bf B154} (1985) 397.}

\lref\atiyah{M.~F.~Atiyah and N.~J.~Hitchin, Phys. Lett. {\bf A107}
(1985) 21.}

\lref\atiyahbook{M.~F.~Atiyah and N.~J.~Hitchin, {\it The Geometry and
Dynamics of Magnetic Monopoles}, Princeton University Press, 1988.}

\lref\strom{A.~Strominger, Nucl. Phys. {\bf B343} (1990) 167.}

\lref\gsw{M.~B.~Green, J.~H.~Schwartz and E.~Witten,
{\it Superstring Theory} vol. 1, Cambridge University Press (1987).}

\lref\polch{J.~Polchinski, Phys. Lett. {\bf B209} (1988) 252.}

\lref\dfluone{M.~J.~Duff and J.~X.~Lu, Nucl. Phys. {\bf B354} (1991) 141.}

\lref\dflutwo{M.~J.~Duff and J.~X.~Lu, Nucl. Phys. {\bf B354} (1991) 129.}

\lref\dfluthree{M.~J.~Duff and J.~X.~Lu, Phys. Rev. Lett. {\bf 66}
(1991) 1402.}

\lref\dflufour{M.~J.~Duff and J.~X.~Lu, Nucl. Phys. {\bf B357} (1991)
534.}

\lref\dfstel{M.~J.~Duff and K.~S.~Stelle, Phys. Lett. {\bf B253} (1991)
113.}

\lref\ferone{R.~C.~Ferrell and D.~M.~Eardley, Phys. Rev. Lett. {\bf 59}
(1987) 1617.}

\lref\fertwo{R.~C.~Ferrell and D.~M.~Eardley, {\it Slowly Moving
Maximally Charged Black Holes} in Frontiers in Numerical Relativity,
Cambridge University Press, 1987.}

\lref\gh{G.~W.~Gibbons and S.~W.~Hawking, Phys. Rev. {\bf D15}
(1977) 2752.}

\lref\ghp{G.~W.~Gibbons, S.~W.~Hawking and M.~J.~Perry, Nucl. Phys. {\bf B318}
(1978) 141.}

\lref\briho{D.~Brill and G.~T.~Horowitz, Phys. Lett. {\bf B262} (1991)
437.}

\lref\gidone{S.~B.~Giddings and A.~Strominger, Nucl. Phys. {\bf B306}
(1988) 890.}

\lref\gidtwo{S.~B.~Giddings and A.~Strominger, Phys. Lett. {\bf B230}
(1989) 46.}

\lref\raj{R.~Rajaraman, {\it Solitons and Instantons}, North Holland,
1982.}

\lref\chsw{P.~Candelas, G.~T.~Horowitz, A.~Strominger and E.~Witten,
Nucl. Phys. {\bf B258} (1984) 46.}

\lref\bogo{E.~B.~Bogomolnyi, Sov. J. Nucl. Phys. {\bf 24} (1976) 449.}

\lref\cogo{E.~Corrigan and P.~Goddard, Comm. Math. Phys. {\bf 80} (1981)
575.}

\lref\wardone{R.~S.~Ward, Comm. Math. Phys. {\bf 79} (1981) 317.}

\lref\wardtwo{R.~S.~Ward, Comm. Math. Phys. {\bf 80} (1981) 563.}

\lref\wardthree{R.~S.~Ward, Phys. Lett. {\bf B158} (1985) 424.}

\lref\groper{D.~J.~Gross and M.~J.~Perry, Nucl. Phys. {\bf B226} (1983)
29.}

\lref\ash{{\it New Perspectives in Canonical Gravity}, ed. A.~Ashtekar,
Bibliopolis, 1988.}

\lref\lich{A.~Lichnerowicz, {\it Th\' eories Relativistes de la
Gravitation et de l'Electro-magnetisme}, (Masson, Paris 1955).}

\lref\goldstein{H.~Goldstein, {\it Classical Mechanics}, Addison-Wesley,
1981.}

\lref\dflufive{M.~J.~Duff and J.~X.~Lu, Class. Quant. Grav. {\bf 9}
(1992) 1.}

\lref\dflusix{M.~J.~Duff and J.~X.~Lu, Phys. Lett. {\bf B273} (1991)
409.}

\lref\hlp{J.~Hughes, J.~Liu and J.~Polchinski, Phys. Lett. {\bf B180}
(1986).}

\lref\town{P.~K.~Townsend, Phys. Lett. {\bf B202} (1988) 53.}

\lref\duff{M.~J.~Duff, Class. Quant. Grav. {\bf 5} (1988).}

\lref\rossi{P.~Rossi, Physics Reports, 86(6) 317-362.}

\lref\gksone{B.~Grossman, T.~W.~Kephart and J.~D.~Stasheff, Commun. Math.
Phys. {\bf 96} (1984) 431.}

\lref\gkstwo{B.~Grossman, T.~W.~Kephart and J.~D.~Stasheff, Commun. Math.
Phys. {\bf 100} (1985) 311.}

\lref\gksthree{B.~Grossman, T.~W.~Kephart and J.~D.~Stasheff, Phys. Lett.
{\bf B220} (1989) 431.}

\newsec{Introduction}

In recent work classical solitonic solutions of string theory with
higher-membrane structure have been investigated. In this paper we consider
some exact bosonic as well as heterotic solutions.

We begin with a review of the results of \rkinst, in which the tree-level
axionic instanton of \reyone\ is extended to an exact solution of bosonic
string theory for the special case of a linear dilaton
wormhole\refs{\abenone,\abentwo}. Exactness is shown by combining
the metric and antisymmetric tensor in a generalized curvature, which
is written covariantly in terms of the tree-level dilaton field, and rescaling
the dilaton order by order in the parameter $\alpha'$\rkinst. The corresponding
conformal field theory is written down.

An exact multi-soliton solution of heterotic string theory\quartet\ with YM
instanton structure in
the four dimensional transverse space can be obtained\refs{\chsone,\chstwo} by
equating the curvature of the Yang-Mills gauge field with the generalized
curvature derived in \rkinst. This solution represents an exact extension of
the tree-level fivebrane solutions of \refs{\strom,\dfluone,\dflutwo}\ and
combines the gauge and axionic instanton structures.

Finally, we consider the recently constructed exact multimonopole solution of
heterotic string theory\refs{\rkmono,\rkmonex}. An interesting aspect
of this string monopole solution is that the divergences stemming
from the YM sector are precisely cancelled by those coming from the gravity
sector, thus resulting in a finite action solution.

For both classes of heterotic solutions, the interplay between the
gravitational and gauge sectors reveals an interesting aspect of
string theory as a finite theory of quantum gravity.

\newsec{Exact Bosonic Solution}

In \reyone\ a tree-level bosonic axionic instanton solution was written
down. Here we review the exact extension of this solution\rkinst\ for the
special case of a wormhole, and write down the corresponding conformal field
theory. For this purpose we use the theorem of equivalence of the
massless string field equations to the sigma-model Weyl invariance
conditions (demonstrated to two-loop order by Metsaev and
Tseytlin\refs{\mtone,\mttwo}), which require the Weyl
anomaly coefficients $\wbg$, $\wbb$ and $\wbd$ to vanish identically to
the appropriate order in the parameter $\alpha'$.
The two-loop solution obtained by this method suggests a
representation of the sigma model as the product of a WZW\gepwit\ model
and a one-dimensional CFT (a Feigin-Fuchs Coulomb gas)\reyone.
This representation allows us to obtain an exact solution.

The bosonic sigma model action can be written as\love
\eqn\sigmod{I={1\over 4\pi\alpha'}\int d^2x\left(\sqrt{\gamma}\gamma^{ab}
\partial_ax^\mu\partial_bx^\nu\met+i\epsilon^{ab}\partial_ax^\mu\partial_b
x^\nu B_{\mu\nu}+\alpha'\sqrt{\gamma}R^{(2)}\phi\right),}
where $\met$ is the sigma model metric, $\phi$ is the dilaton and $B_{\mu\nu}$
is the antisymmetric tensor and where $\gamma_{ab}$ is the worldsheet metric
and $R^{(2)}$ the two-dimensional curvature.
The Weyl anomaly coefficients are given by\refs{\mtone,\mttwo}
\eqn\weycos{\eqalign{\wbg&=\alpha'(\hat R_{(\mu\nu)}+2\nabla_\mu
\nabla_\nu\phi)\cr&+{\alpha'^2\over
2}\left(\hat R^{\alpha\beta\gamma}{}_{(\mu}\hat R_{\nu)\alpha\beta\gamma}
-{1\over 2}\hat R^{\beta\gamma\alpha}{}_{(\mu}\hat
R_{\nu)\alpha\beta\gamma}+{1\over 2}\hat R_{\alpha(\mu\nu)\beta}
(H^2)^{\alpha\beta}\right)+\nabla_{(\mu}W_{\nu)},\cr
\wbb&=\alpha'(\hat R_{[\mu\nu]}+H_{\mu\nu}{}^\lambda\partial_\lambda\phi)
\cr&+{\alpha'^2\over
2}\left(\hat R^{\alpha\beta\gamma}{}_{[\mu}\hat R_{\nu]\alpha\beta\gamma}
-{1\over 2}\hat R^{\beta\gamma\alpha}{}_{[\mu}\hat
R_{\nu]\alpha\beta\gamma}+{1\over 2}\hat R_{\alpha[\mu\nu]\beta}
(H^2)^{\alpha\beta}\right)+{1\over 2}H_{\mu\nu}{}^\lambda W_\lambda
,\cr
\wbd&={D\over 6}-{\alpha'\over 2}\left(\nabla^2\phi-2(\partial\phi)^2+{1\over
12}H^2\right)\cr&+{\alpha'^2\over 16}\left(2(H^2)^{\mu\nu}\nabla_\mu
\nabla_\nu\phi+R^2_{\lambda\mu\nu\rho}-{11\over 2}RHH
+{5\over 24}H^4+{11\over 8}(H^2_{\mu\nu})^2+{4\over 3}\nabla H\cdot
\nabla H\right)\cr&+{1\over 2}W^\lambda\partial_\lambda\phi,\cr}}
where $H_{\mu\nu\lambda}=\partial_{[\mu}B_{\nu\lambda]}$,
$W_\mu=-(\alpha'^2/24)\nabla_\mu H^2$,
$\nabla H\cdot\nabla H\equiv\nabla_\alpha H_{\beta\gamma\delta}
\nabla^\alpha H^{\beta\gamma\delta}$
and $\grijkl$ is the generalized curvature defined in terms of the standard
curvature $\rijkl$ and $H_{\mu\alpha\beta}$ by\fridven\
\eqn\gcurv{\grijkl=\rijkl+{1\over
2}\left(\nabla_lH^i{}_{jk}-\nabla_kH^i{}_{jl}\right)
+{1\over 4}\left(H^m{}_{jk}H^i{}_{lm}-H^m{}_{jl}H^i{}_{km}\right).}
One can also define $\grijkl$ as the Riemann tensor generated
by the generalized Christoffel symbols $\hat\Gamma^\mu_{\alpha\beta}$
where  $\hat\Gamma^\mu_{\alpha\beta}=\Gamma^\mu_{\alpha\beta}
-(1/2) H^\mu{}_{\alpha\beta}$.
Unless otherwise indicated, all expressions are written to two loop
order in the beta-functions, which corresponds to $O(\alpha')$ in the action.

For any dilaton function satisfying $e^{-2\phi}\Box\ e^{2\phi}=0$ with
\eqn\sansatz{\eqalign{\met&=e^{2\phi}\delta_{\mu\nu}\qquad \mu,\nu=1,2,3,4,\cr
g_{ab}&=\delta_{ab}\qquad\quad   a,b=5,...,26,\cr
H_{\mu\nu\lambda}&=\pm\epsilon_{\mu\nu\lambda\sigma}\partial^\sigma\phi
\qquad \mu,\nu,\lambda,\sigma=1,2,3,4\cr}}
the $O(\alpha')$ Weyl anomaly coefficients vanish\rkinst. This can be
seen as follows. For any solution of the form \sansatz,
we can express the generalized curvature in covariant form in terms of
the dilaton field $\phi$:
\eqn\gcurvphi{\grijkl=\delta_{il}\nabla_k\nabla_j\phi
-\delta_{ik}\nabla_l\nabla_j\phi+\delta_{jk}\nabla_l\nabla_i\phi
-\delta_{jl}\nabla_k\nabla_i\phi\pm\epsilon_{ijkm}\nabla_l\nabla_m\phi
\mp\epsilon_{ijlm}\nabla_k\nabla_m\phi.}
It follows from \gcurvphi\ that
\eqn\gricci{\eqalign{\hat R_{(\mu\nu)}&=-2\nabla_\mu\nabla_\nu\phi,\cr
\hat R_{[\mu\nu]}&=0.\cr}}
It also follows from \sansatz\ that
\eqn\zeroterms{\eqalign{\nabla^2\phi&=0,\cr{H_{\mu\nu}}^\lambda
\partial_\lambda\phi&=0,\cr H^2&=24(\partial\phi)^2.\cr}}
{}From \gricci\ and \zeroterms\ it follows that the $O(\alpha')$ terms
in the Weyl anomaly coefficients in \weycos\ vanish identically for the
ansatz \sansatz. A tree-level multi-instanton solution is therefore
given by \sansatz\ with the dilaton given by
\eqn\multy{e^{2\phi}=C+\sum_{i=1}^N {Q_i\over |\vec x -\vec a_i|^2},}
where $Q_i$ is the charge and $\vec a_i$ the location in the four-space
$(1234)$ of the $i$th instanton. We call $(1234)$ the transverse space,
as the solitons have the structure of $21+1$-dimensional objects embedded
in a $26$-dimensional spacetime.

We now specialize to the spherically symmetric case of $e^{2\phi}={Q/r^2}$ in
\sansatz\ and determine the $O(\alpha')$ corrections to the massless fields in
\sansatz\ so that the Weyl anomaly coefficients vanish to $O(\alpha'^2)$.
For this solution we notice
\eqn\ddphi{\nabla_\mu\nabla_\nu\phi=0,}
and therefore from \gcurvphi
\eqn\gcurvzero{\grijkl=0,}
and we have what is called a ``parallelizable" space\refs{\mtone,\mttwo}.
To maintain a parallelizable space to $O(\alpha')$ we keep $\met$
and $H_{\alpha\beta\gamma}$ in their lowest order form and assume
that any corrections to \sansatz\ appear in the dilaton:
\eqn\tlgansatz{\eqalign{\phi&=\phi_0+\alpha'\phi_1+...\cr
e^{2\phi_0}&={Q\over r^2},\cr
\met&=e^{2\phi_0}\delta_{\mu\nu},\cr
H_{\mu\nu\lambda}&=\pm\epsilon_{\mu\nu\lambda\sigma}\partial^\sigma\phi_0
.\cr}}
It follows from \tlgansatz\ that $H^2=24(\partial\phi_0)^2=24/Q$ and
thus $W_\mu=0$. It follows from \gcurvzero\ that $\wbg$ and $\wbb$
vanish identically to two loop order and that
\eqn\dweylone{\eqalign{\wbd={D\over 6}+&\alpha'\left((\partial\phi)^2
-{1\over Q} \right)\cr
&+{\alpha'^2\over 16}\left(R^2_{\lambda\mu\nu\rho}-{11\over 2}RHH
+{5\over 24}H^4+{11\over 8}(H^2_{\mu\nu})^2+{4\over 3}\nabla H\cdot\nabla H
\right).\cr}}
We use the relations in equation (34) in \mtone\ for parallelizable spaces
and the observation that $(H^2_{\mu\nu})^2=2H^4=192/Q^2$ for our solution to
get the identities
\eqn\parallel{\eqalign{R^2_{\lambda\mu\nu\rho}&={1\over 8}H^4,\cr
RHH&={1\over 2}H^4,\cr\nabla H\cdot\nabla H&=0 .\cr}}
\dweylone\ then simplifies further to
\eqn\dweyltwo{\wbd={D\over 6}+\alpha'\left((\partial\phi)^2-{1\over Q}\right)
+2{\alpha'^2\over Q^2}.}
The lowest order term in $\wbd$ is proportional to the central charge
and the $O(\alpha')$ terms vanish identically. With the choice
$\overrightarrow\nabla\phi_1=-(1/Q)\overrightarrow\nabla\phi_0$, the
$O(\alpha'^2)$ terms also vanish identically.
The two-loop solution is then given by
\eqn\tlsansatz{\eqalign{e^{2\phi}&={Q\over r^{2(1-{\alpha'\over Q})}},\cr
\met&={Q\over r^2}\delta_{\mu\nu},\cr
H_{\mu\nu\lambda}&=\pm\epsilon_{\mu\nu\lambda\sigma}\partial^\sigma\phi_0,\cr}}
which corresponds to a simple rescaling of the dilaton.
A quick check shows that this solution has finite action near the
singularity.

We now rewrite $\wbd$ in \dweyltwo\ in the following suggestive form:
\eqn\dweylsplit{\eqalign{6\wbd&=\left(1+6\alpha'(\partial\phi)^2\right)
+\left(3-6{\alpha'\over Q}+12({\alpha'\over Q})^2\right)\cr&=4.\cr}}
The above splitting of the central charge $c=6\wbd$ suggests
the decomposition of the corresponding sigma model into the
product of a one-dimensional CFT (a Feigin-Fuchs Coulomb gas)
and a three-dimensional WZW model with an
$SU(2)$ group manifold \refs{\reyone,\mtone,\mttwo}.
This can be seen as follows.
Setting $u=\ln r$, we can rewrite \sigmod\ for our solution\reyone\
in the form $I=I_1+I_3$, where
\eqn\onecft{I_1={1\over 4\pi\alpha'}\int d^2x\left(Q(\partial u)^2
+\alpha' R^{(2)}\phi\right)}
is the action for a Feigin-Fuchs Coulomb gas, which is a one-dimensional
CFT with central charge given by
$c_1=1+6\alpha'(\partial\phi)^2$\ginsp. The imaginary charge of the
Feigin-Fuchs Coulomb gas describes the dilaton background growing
linearly in imaginary time\refs{\abenone,\abentwo}.
$I_3$ is the Wess--Zumino--Witten\gepwit\ action on an $SU(2)$ group manifold
with central charge
\eqn\threecharge{c_3={3k\over k+2}\simeq 3-{6\over k}+{12\over k^2}+...}
where $k=Q/\alpha'$, called the ``level" of the WZW model, is an
integer. This can be seen from
the quantization condition on the Wess-Zumino term\gepwit
\eqn\iwzw{\eqalign{I_{WZ}&={i\over 4\pi\alpha'}\int_{\partial S_3}
d^2x\epsilon^{ab}\partial_ax^\mu\partial_bx^\nu B_{\mu\nu}\cr
&={i\over 12\pi\alpha'}\int_{S_3}d^3x\epsilon^{abc}
\partial_ax^\mu \partial_bx^\nu\partial_cx^\lambda H_{\mu\nu\lambda}\cr
&=2\pi i\left({Q\over\alpha'}\right).\cr}}
Thus $Q$ is not arbitrary, but is quantized in units of $\alpha'$.

We use this splitting to obtain exact expressions
for the fields by fixing the metric and antisymmetric tensor field
in their lowest order form and rescaling the dilaton order by order in
$\alpha'$. The resulting expression for the dilaton is
\eqn\alldilaton{e^{2\phi}={Q\over r^{\sqrt{{4\over 1+{2\alpha'\over Q}}}}}.}

\newsec{Exact Heterotic Multi-Instanton Solution}

We now turn to the heterotic multi-instanton solution of
\refs{\chsone,\chstwo}.
The tree-level supersymmetric vacuum equations for the heterotic string
are given by
\eqn\suei{\delta\psi_M=\left(\nabla_M-{\textstyle {1\over 4}}H_{MAB}\Gamma^{AB}
\right)\epsilon=0,}
\eqn\sueii{\delta\lambda=\left(\Gamma^A\partial_A\phi-{\textstyle{1\over 6}}
H_{AMC}\Gamma^{ABC}\right)\epsilon=0,}
\eqn\sueiii{\delta\chi=F_{AB}\Gamma^{AB}\epsilon=0,}
where $\psi_M,\ \lambda$ and $\chi$ are the gravitino, dilatino and gaugino
fields. The Bianchi identity is given by
\eqn\bianchi{dH=\alpha' \left(\tr R\wedge R-{\textstyle{1\over 30}}\Tr
        F\wedge F\right).}

The $(9+1)$-dimensional Majorana-Weyl fermions decompose down to
chiral spinors according to $SO(9,1)\supset SO(5,1) \otimes SO(4)$ for
the $M^{9,1}\to M^{5,1}\times M^4$ decomposition.
Let $\mu,\nu,\lambda,\sigma=1,2,3,4$ and $a,b=0,5,6,7,8,9$. Then the ansatz
\eqn\anstz{\eqalign{\met&=e^{2\phi}\delta_{\mu\nu},\cr g_{ab}&=\eta_{ab},\cr
H_{\mu\nu\lambda}&=\pm\epsilon_{\mu\nu\lambda\sigma}\partial^\sigma\phi\cr}}
with constant chiral spinors $\epsilon_\pm$ solves the supersymmetry
equations with zero background fermi fields provided the YM gauge
field satisfies the instanton (anti)self-duality condition
\eqn\yminst{F_{\mu\nu}=\pm {1\over 2}\epsilon_{\mu\nu}{}^{\lambda\sigma}
F_{\lambda\sigma}.}
An exact solution is obtained as follows. Define a generalized connection by
\eqn\genc{\Omega^{AB}_{\pm M}=\omega^{AB}_M\pm H^{AB}_M}
embedded in an SU(2) subgroup of the gauge group, and equate it
to the gauge connection $A_\mu$\dine\ so that $dH=0$ and the corresponding
curvature $R(\Omega_{\pm})$ cancels against the Yang-Mills field strength $F$.
As in the bosonic case, for $e^{-2\phi}\Box\ e^{2\phi}=0$ with the
above ansatz, the curvature of the generalized connection can be written in the
covariant form\rkinst
\eqn\gencurv{\eqalign{R(\Omega_\pm)_{\mu\nu}^{mn}
=&\delta_{n\nu}\nabla_m\nabla_\mu\phi
- \delta_{n\mu}\nabla_m\nabla_\nu\phi + \delta_{m\mu}\nabla_n\nabla_\nu\phi
- \delta_{m\nu}\nabla_n\nabla_\mu\phi \cr
&\pm \epsilon_{\mu mn\alpha}\nabla_\alpha\nabla_\nu\phi
\mp \epsilon_{\nu mn\alpha}\nabla_\alpha\nabla_\mu\phi ,\cr}}
from which it easily follows that
\eqn\gcinst{R(\Omega_\pm)^{mn}_{\mu\nu}=
\mp\half\epsilon_{\mu\nu}^{\ \ \ \lambda\sigma}
R(\Omega_{\pm})_{\lambda\sigma}^{mn}.}
Thus we have a solution with the ansatz \anstz\ such that
\eqn\exsol{F_{\mu\nu}^{mn}=R(\Omega_{\pm})_{\mu\nu}^{mn},}
where both $F$ and $R$ are (anti)self-dual.
This solution becomes exact since $A_\mu=\Omega_{\pm\mu}$
implies that all the higher order corrections
vanish\refs{\dine,\brone,\brtwo,\chsone,\chstwo,\rkdg}.
The self-dual solution for the gauge connection is then given by the 't Hooft
ansatz
\eqn\hfanstz{A_\mu=i \overline{\Sigma}_{\mu\nu}\partial_\nu \ln f.}
For a multi-instanton solution $f$ is given by
\eqn\finst{f=e^{-2\phi_0}e^{2\phi}
=1+\sum_{i=1}^N{\rho_i^2\over |\vec x - \vec a_i|^2},}
where $\rho_i^2$ is the instanton scale size and $\vec a_i$ the location in
four-space of the $i$th instanton. An interesting feauture of the heterotic
solution is that it combines a YM instanton structure in the gauge
sector with an axionic instanton structure in the gravity sector.
In addition, the heterotic solution has finite action.

Note that the single instanton solution in the heterotic case
carries through to higher order without correction to the dilaton.
This seems to contradict the bosonic solution by suggesting that
the expansion for the Weyl anomaly coefficient $\wbd$ terminates at one loop.
This contradiction is resolved by noting that for a
supersymmetric ansatz the bosonic contribution to the
central charge is given by\swzw
\eqn\scharge{c_3={3k'\over k'+2}~,}
where $k'=k-2$. This reduces to
\eqn\termcharge{\eqalign{c_3&=3-{6\over k}\cr &=3-{6\alpha'\over Q},\cr}}
which indeed terminates at one loop order. The exactness of the splitting
then requires that $c_1$ not get any corrections from
$(\partial\Phi)^2$ so that $c_1+c_3=4$ is exact for the tree-level value
of the dilaton\refs{\chsone,\chstwo,\rkdg}.

The exact heterotic instanton represents a nonperturbative classical
solution combining the massless fields in the string gravitational sector
with the YM field in the gauge sector. If the extension of this solution
to an exact string loop solution is also nonperturbative, the heterotic
instanton may provide a potential testing ground for string theory as a finite
theory of quantum gravity. In addition, there is the hope that this type of
finite action instanton solution may eventually lead to an understanding of the
vacuum in string theory, a role analogous to that of the instanton in YM field
theory.

\newsec{Exact Heterotic Multimonopole Solution}

In this section we consider the exact multimonopole solution of heterotic
string theory
obtained in \refs{\rkmono,\rkmonex}. The derivation of this solution closely
parallels that of the multi-instanton solution reviewed in section 3, but
in this case, the solution possesses three-dimensional (rather than
four-dimensional) spherical symmetry near each source. The reduction is
effected by singling out a direction in the transverse space. An exact solution
is now given by
\eqn\anstz{\eqalign{\met&=e^{2\phi}\delta_{\mu\nu},\qquad g_{ab}=\eta_{ab},\cr
H_{\mu\nu\lambda}&=\pm\epsilon_{\mu\nu\lambda\sigma}\partial^\sigma\phi,\cr
e^{2\phi}&=e^{2\phi_0}f,\cr
A_\mu&=i \overline{\Sigma}_{\mu\nu}\partial_\nu \ln f,\cr}}
where in this case
\eqn\fdmono{f=1+\sum_{i=1}^N{m_i\over |\vec x - \vec a_i|},}
where $m_i$ is the charge and $\vec a_i$ the location in
the three-space $(123)$ of the $i$th monopole. If we identify the
scalar field as $\Phi\equiv A_4$, then the $SU(2)$ gauge and scalar fields may
be simply written in terms of the dilaton as\refs{\rkmono,\rkmonex}
\eqn\stmono{\eqalign{\Phi^a&=-{2\over g}\delta^{ia}\partial_i\phi,\cr
A_k^a&=-{2\over g}\epsilon^{akj}\partial_j\phi\cr}}
for the self-dual solution. For the anti-self-dual solution, the scalar
field simply changes sign. Here $g$ is the YM coupling constant. Note
that $\phi_0$ drops out in \stmono.

The above solution (with the gravitational fields obtained
directly from \anstz\ and \fdmono) represents an exact multimonopole
solution of heterotic string theory and has the same structure in
the four-dimensional transverse space as an analogous multimonopole solution
of the YM + scalar field action\refs{\rkmono,\rkmonex}. If we identify
the $(123)$ subspace of the transverse space as the space part of the
four-dimensional spacetime (with some toroidal compactification, similar
to that used in \jim) and take the timelike direction as the usual $X^0$,
then the monopole properties of the field theory solution carry
over directly into the string solution.

The string action contains a term $-\alpha' F^2$ which diverges
as in the field theory solution (see \rkmono). However, this divergence
is precisely cancelled by the term $\alpha' R^2(\Omega_\pm)$ in the
$O(\alpha')$
action. This result follows from the exactness condition
$A_\mu=\Omega_{\pm\mu}$
which leads to $dH=0$ and the vanishing of all higher order corrections
in $\alpha'$. Another way of seeing this is to consider the higher order
corrections to the bosonic action shown in \refs{\brone,\brtwo}. All such
terms contain the tensor $T_{MNPQ}$, a generalized curvature incorporating
both $R(\Omega_\pm)$ and $F$. The ansatz is constructed precisely so that this
tensor vanishes identically\refs{\rkinst,\rkdg}. The action thus reduces to
its finite lowest order form and can be calculated directly for a multi-source
solution from the expressions for the massless fields in the gravity sector.

The divergences in the gravitational sector in heterotic string theory thus
serve to cancel the divergences stemming from the field theory solution. This
solution thus provides an interesting example of how this type of cancellation
can occur in string theory, and supports the promise of string theory as a
finite theory of quantum gravity. Another point of interest is that the string
solution represents a supersymmetric multimonopole solution coupled to gravity,
whose zero-force condition in the gravity sector (cancellation of
the attractive gravitational force and repulsive antisymmetric field force)
arises as a direct result of the zero-force condition in the gauge sector
(cancellation of vector and scalar forces of exchange) once the gauge
connection
and generalized connection are identified.

\newsec{Conclusion}

We outlined in section 2 the bosonic tree-level axionic instanton solution of
\reyone\ and its exact extension for the case of a single instanton wormhole
solution\rkinst. A combination of the YM gauge instanton and the axionic
instanton solution led to an exact multi-instanton solution in heterotic string
theory\refs{\chsone,\chstwo\rkdg}, which was discussed in section 3.
Finally, in section 4 we turned to the recently constructed exact multimonopole
solution of heterotic string theory\refs{\rkmono,\rkmonex}.

For both heterotic instantons and monopoles, if the nonperturbative nature
of these classical solutions carries over to their quantum string-loop
extensions, we may eventually gain some insight into the nature of string
theory
as a finite theory of quantum gravity. The heterotic instanton solution may
perhaps play a role analogous to that of instantons in field theory by
providing an understanding of the structure of the vacuum in string theory.

An interesting aspect of the monopole solution is that the YM divergences
of the modified 't Hooft ansatz are precisely cancelled in
the string theory solution by similar divergences in the gravity sector,
resulting in a finite action solution. This finding is
significant in that it represents an example of how string theory
incorporates gravity in such a way as to cancel infinities inherent in
gauge theories. It will be especially noteworthy if the quantum string loop
extension of this solution retains this feature.


\vfil\eject
\listrefs
\bye